\documentclass{aa} 
\usepackage{graphicx}
\usepackage{txfonts}
\usepackage{ulem}
\usepackage{natbib,twoopt}
\usepackage[breaklinks,urlbordercolor={1 1 0}]{hyperref}  
\bibpunct{(}{)}{;}{a}{}{,} 
\newcommandtwoopt{\citeads}[3][][]{\href{http://adsabs.harvard.edu/abs/#3}%
{\citealp[#1][#2]{#3}}}
\newcommandtwoopt{\citepads}[3][][]{\href{http://adsabs.harvard.edu/abs/#3}%
{\citep[#1][#2]{#3}}}
\newcommandtwoopt{\citetads}[3][][]{\href{http://adsabs.harvard.edu/abs/#3}%
{\citet[#1][#2]{#3}}} 
\newcommandtwoopt{\citeyearads}[3][][]%
{\href{http://adsabs.harvard.edu/abs/#3}{\citeyear[#1][#2]{#3}}} 
\def\hh{H$_2$}
\def\emin{e$^-$}
\def\cp{C$^+$}
\def\op{O$^+$}
\def\ohp{OH$^+$}
\def\chp{CH$^+$}
\def\shp{SH$^+$}
\def\cop{CO$^+$}
\def\thco{$^{13}$CO}
\def\hp{H$^+$}
\def\hhhp{H$_3^+$}
\def\hho{H$_2$O}
\def\hhop{H$_2$O$^+$}
\def\hhhop{H$_3$O$^+$}
\def\hnop{H$_n$O$^+$}
\def\nn{N$_2$}
\def\nnhp{N$_2$H$^+$}
\def\hcop{HCO$^+$}
\def\ps{s$^{-1}$}
\def\mic{$\mu$m}
\def\kms{km\,s$^{-1}$}
\def\kkms{K\,km\,s$^{-1}$}
\def\pow#1#2{#1$\times$10$^{#2}$}
\def\scm{cm$^{-2}$}
\def\ccm{cm$^{-3}$}
\def\ccps{cm$^3$\,s$^{-1}$}

\def\dv{$\Delta${\it V}}
\def\vlsr{$V_{\rm LSR}$}
\def\gtsim{{_>\atop{^\sim}}}
\def\ltsim{{_<\atop{^\sim}}}

\def\tkin{$T_{\rm kin}$}

\def\txc{$T_{\rm ex}$}
\def\tas{$T_A^*$}

\def\new#1{{#1}}
\def\newer#1{{#1}}
\def\newest#1{{#1}}
\def\old{}
\def\oldest{}
\begin{document}
\title{Spatially extended \ohp\ emission \new{from the Orion Bar and Ridge}.
\thanks{\textit{Herschel} is an ESA space observatory with science instruments provided
by European-led Principal Investigator consortia and with important participation from NASA}}
\titlerunning{\ohp\ emission in Orion.}

\author{F.F.S. van der Tak \inst{\ref{sron},\ref{rug}} \and
              Z. Nagy \inst{\ref{rug},\ref{sron}} \and
              V. Ossenkopf \inst{\ref{cologne}} \and
              Z. Makai \inst{\ref{cologne}} \and
              J.H. Black \inst{\ref{onsala}} \and
              A. Faure \inst{\ref{laog}} \and
              M. Gerin \inst{\ref{lerma}} \and
              E.A. Bergin \inst{\ref{mich}} 
              }
              
\institute{SRON Netherlands Institute for Space Research, Landleven 12, 9747 AD Groningen, The Netherlands; \email{vdtak@sron.nl} \label{sron} \and
          Kapteyn Astronomical Institute, University of Groningen, The Netherlands \label{rug} \and
          I. Physikalisches Institut, Universit\"at zu K\"oln, Germany \label{cologne} \and
          Chalmers University of Technology, Onsala Space Observatory, Sweden \label{onsala} \and
          UJF Grenoble, IPAG, France \label{laog} \and
          LERMA, CNRS, Observatoire de Paris and ENS, France \label{lerma} \and
          Department of Astronomy, University of Michigan, USA \label{mich}
          }
          
\date{Submitted June 28, 2013; accepted November 8, 2013}

\abstract
{The reactive \hnop\ ions (\ohp, \hhop\ and \hhhop) are widespread in the interstellar medium and act as precursors to the \hho\ molecule. While \hnop\ absorption is seen on many Galactic lines of sight, active galactic nuclei often show the lines in emission.}
{This paper shows the first example of a Galactic source of \hnop\ line emission: the Orion Bar, a bright nearby photon-dominated region (PDR).}
{We present line profiles and maps of \ohp\ line emission toward the Orion Bar, and upper limits to \hhop\ and \hhhop\ lines. 
We analyze these HIFI data with non-LTE radiative transfer and PDR chemical models, using newly calculated inelastic collision data for the e-\ohp\ system.}
{Line emission is detected over $\sim$1$'$ (0.12~pc), tracing the Bar itself as well as a perpendicular feature identified as the Southern tip of the Orion Ridge, which borders the Orion Nebula on its Western side. 
The line width of $\approx$4~\kms\ suggests an origin of the \ohp\ emission close to the PDR surface, at a depth of $A_V \sim$0.3--0.5 into the cloud \new{where most hydrogen is in atomic form}. 
Models with collisional and radiative excitation of \ohp\ require unrealistically high column densities to match the observed line intensity, indicating that \newest{the formation of \ohp\ in the Bar} is rapid enough to influence its excitation.
Our best-fit \ohp\ column density of  $\sim$\pow{1.0}{14}~\scm\ is similar to that in previous absorption line studies, while our limits on the ratios of \ohp/\hhop\ \new{($\gtsim$40)} and \ohp/\hhhop\ \new{($\gtsim$15)} are somewhat higher than seen before. }
{The column density of \ohp\ is consistent with estimates from \newest{a thermo-chemical model} for parameters applicable to the Orion Bar, given the current uncertainties in the local gas pressure and the spectral shape of the ionizing radiation field.
The unusually high \ohp/\hhop\ and \ohp/\hhhop\ ratios are probably due to the high UV radiation field and electron density in this object. 
Photodissociation and electron recombination are more effective destroyers of \ohp\  than the reaction with \hh, which limits the production of \hhop. 
The appearance of the \ohp\ lines in emission is the result of the high \newer{density of electrons and H atoms in the Orion Bar, since for these species, inelastic collisions with \ohp\ are faster than reactive ones}.
In addition, chemical pumping, far-infrared pumping by local dust, and near-UV pumping by Trapezium starlight contribute to the \ohp\ excitation.
Similar conditions may apply to extragalactic nuclei where \hnop\ lines are seen in emission.}


\maketitle

\section{Introduction}
\label{s:intro}

Although interstellar clouds have ionization fractions of only 10$^{-4}$--10$^{-8}$, ionic species are very useful to probe physical conditions in such clouds \citepads{2012RPPh...75f6901L}. 
In diffuse clouds ($A_V < 1$), the main ion source is UV photoionization of carbon, while in dense clouds ($A_V > 1$), cosmic-ray ionization of hydrogen is the dominant ionization mechanism \citepads{2007ARA&A..45..339B}. 
Proton transfer reactions of interstellar \hhhp\ with abundant species such as CO and \nn\ lead to \hcop\ and \nnhp, which are widely observed in the interstellar medium. 
Such stable ionic species are useful as tracers of the interaction of interstellar gas with magnetic fields 
(Houde et al \citeyearads{2004ApJ...604..717H}, Schmid-Burgk et al \citeyearads{2004A&A...419..949S}), whereas ions which react rapidly with \hh\ trace other parameters such as the gas density and the ionization rate. 

At temperatures $\ltsim$250~K, the formation of interstellar \hho\ in the gas phase proceeds through a series of ion-molecule reactions. 
After charge transfer of \hp\ or \hhhp\ to O, repeated reactions of \op\ with \hh\ produce \ohp, \hhop, and finally \hhhop, which upon dissociative recombination with a free electron produces \hho. 
The \ohp\ and \hhop\ ions are well known from the spectra of comets where they appear as photodissociation products of \hho\ \citepads{1950ApJ...111..530S,1974A&A....31..123H}.
In the interstellar medium, \hho\ and \hhhop\ have been known for decades \citepads{1992ApJ...399..533P}, but observation of the intermediate products \ohp\ and \hhop\ had to await the launch of ESA's {\it Herschel} Space Observatory \citepads{2010A&A...518L...1P}. 
Strong absorption in rotational lines of interstellar \ohp\ and \hhop\ is seen with Herschel on many lines of sight in our Galaxy (Gerin et al \citeyearads{2010A&A...518L.110G}, Ossenkopf et al \citeyearads{2010A&A...518L.111O}) and even some beyond \citepads{2010A&A...521L...1W}. 
In addition, electronic absorption lines of \ohp\ have been reported in sensitive near-UV spectra of several diffuse interstellar clouds \citepads{2010ApJ...719L..20K}.

The Herschel data, as well as the single \ohp\ line observed from the ground toward the Galactic Center source SgrB2 \citepads{2010A&A...518A..26W}, imply large column densities of \ohp\ and \hhop.
The hydrogen in the absorbing clouds thus cannot be purely in atomic form, because no \ohp\ and \hhop\ would be produced, nor in purely molecular form, because all \ohp\ and \hhop\ would react into \hhhop\ and \hho. 
Using models of UV-irradiated interstellar clouds (PDRs), the observed abundances of \ohp\ and \hhop\ can be used to infer the relative fractions of hydrogen in atomic and molecular forms \citepads{2010A&A...521L..10N}, which itself traces the ionization rates of the clouds \citepads{2012ApJ...754..105H}.

While the interpretation of interstellar \hnop\ absorption is reasonably well understood, lines of \ohp\ and \hhop\ have also been observed in emission from the nuclei of several active galaxies, most famously Mrk~231 \citepads{2010A&A...518L..42V}. 
The large dipole moments and small reduced masses of the \hnop\ ions imply high line frequencies and large radiative decay rates, so that collisional excitation of their rotational levels requires extremely high densities and line emission is not expected to be observable. 
Understanding this phenomenon benefits from finding a Galactic source of \hnop\ line emission, which can be studied in more detail than extragalactic nuclei.

This paper presents the first observation of \ohp\ line emission toward a source within our Galaxy: the Orion Bar. 
Due to its brightness and nearly edge-on geometry, this PDR is well-suited to observe physical and chemical changes in the gas as a function of depth into the cloud, as the intensity of UV irradiation by the Trapezium stars decreases \citepads[e.g.,]{2009A&A...498..161V}.
The Orion Bar is also notable as the only known Galactic source of \new{interstellar} HF line emission \citepads{2012A&A...537L..10V}.
In this case, the proximity of this region (420~pc: Menten et al \citeyearads{2007A&A...474..515M}, Hirota et al \citeyearads{2007PASJ...59..897H}) allows us to resolve the \hnop\ line emission both spatially and spectrally. 
We use non-LTE radiative transfer models and PDR thermo-chemical models to interpret our results.

\section{Observations}
\label{s:obs}

\begin{table}
\caption{Observed lines.}
\label{t:lines}
\begin{tabular}{ccccc}
\hline \hline
\noalign{\smallskip}
\multicolumn{2}{c}{Molecule / Transition}  & Frequency  & $E_{\rm up}$ & $A_{\rm ul}$  \\
                 &                      & GHz             & K                      & s$^{-1}$        \\ 
\noalign{\smallskip}
\hline
\noalign{\smallskip}
\ohp\    & $1_0$--$0_1$ $F$=1/2--3/2 & \phantom{1}909.159  &  43.6  & 0.011 \\
\ohp\    & $1_2$--$0_1$ $F$=5/2--3/2\tablefootmark{a} & \phantom{1}971.804  & 46.7  & 0.033 \\
\ohp\    & $1_1$--$0_1$ $F$=3/2--3/2 &                    1033.119   & 49.6  & 0.018 \\
p-\hhop\    &  $1_{10}$--$1_{01}$ & \phantom{1}607.227 & 59.2 & 0.006 \\
o-\hhop\    &  $1_{11}$--$0_{00}$ &                     1115.186 & 53.6 & 0.027 \\
\hhhop\  & $0_0^-$--$1_0^+$    & \phantom{1}984.709  & 54.7 & 0.023 \\
\hhhop\  & $1_1^-$--$1_1^+$    &                     1655.831  & 79.5 & 0.055 \\
\noalign{\smallskip}
\hline
\noalign{\smallskip}
\end{tabular}
\tablefoot{The strongest hyperfine component is listed, unless otherwise noted. \\
\tablefoottext{a}{Blend with the $F = 3/2 -1/2$ hyperfine component. The frequency and $E_{\rm up}$ are averages, while the $A_{\rm ul}$ is the sum.}}
\end{table}

The CO$^+$ peak ($\alpha_\mathrm{J2000}=\rm{05^h35^m20.6^s}$, $\delta_\mathrm{J2000}=-05^\circ 25'14''$) in the Orion Bar \citepads{1995A&A...296L...9S} has been observed as a spectral scan over the full HIFI range as part of the HEXOS (Herschel observations of EXtra-Ordinary Sources) guaranteed-time key program \citepads{2010A&A...521L..20B} using the HIFI instrument \citepads{2010A&A...518L...6D} of the Herschel Space Observatory \citepads{2010A&A...518L...1P}.
This paper uses data from HIFI bands 1b (H$_2$O$^+$ $1_{10}$--$1_{01}$), 3b (OH$^+$ $1_0$--$0_1$), 4a (OH$^+$ $1_2$--$0_1$ and $1_1$--$0_1$), 5a (H$_2$O$^+$ $1_{11}$--$0_{00}$), and 6b (\hhhop\ $1_1^-$--$1_1^+$).
These observations were carried out in 2011 March and April in load chop mode with a redundancy of 4, except that frequency switching was used in band 5.
On-source integration times are $\approx$50~s for most spectra except \ohp\ $1_0$--$0_1$ (20~s) and \hhop\ $1_{11}$--$0_{00}$ (190~s).
\new{The ObsIDs of the spectra, without the leading 1342, are 215923 for Band 1b, 216380 for Band 3b, 218628 for Band 4a, 194666 for Band 5a and 218426 for Band 6b}.

Table~\ref{t:lines} lists the frequencies of the lines as well as other spectroscopic parameters, which have been taken from the CDMS database \citepads{2001A&A...370L..49M}\footnote{\tt http://www.astro.uni-koeln.de/cdms/}.
The size of the telescope beam for these observations is 19--23$''$ FWHM, corresponding to 9000 AU or 0.04 pc, except for the \hhop\ $1_{10}$--$1_{01}$ line where it is 35$''$, and for the \hhhop\ $1_1^-$--$1_1^+$ line where it is 15$''$.
The WBS (Wide-Band Spectrometer) was used as backend, covering 4 GHz bandwidth in four 1140~MHz subbands at 1.1~MHz resolution. 
The velocity calibration of HIFI data is accurate to $\sim$0.5 kms$^{-1}$ or better. 
The data were reduced with HIPE \citepads{2010ASPC..434..139O} pipeline version 6.0, using the task \textit{doDeconvolution} for the sideband deconvolution, while further analysis was done in the CLASS package.

In addition to the HIFI spectral scans, a $115''\times65''$ area centered on $\alpha$ = 05:35:20.81, $\delta$ = --05:25:17.1 with a position angle of 145$^\circ$ was mapped in the OH$^+$ $1_2-0_1$ transition with HIFI, in On-The-Fly (OTF) mapping mode with position-switch reference, using a total integration time of 20 min. 
\new{The noise level of the map is 0.22~K per 0.7~\kms\ channel and its ObsID is 218216.} 
The \new{fully sampled} map was reduced with HIPE 6.0 and exported to CLASS for further analysis.
\new{We compare these data to a map of the same area in the CO 10-9 line at $\nu$ = 1151.985 GHz ($E_{\rm up}$ = 304~K), also observed within the HEXOS program under ObsID 217736, with a noise level of 0.56~K per 1.0~\kms\ channel.}

\section{Results}
\label{s:res}

\begin{figure}[t]
\centering
\includegraphics[width=9cm,angle=0]{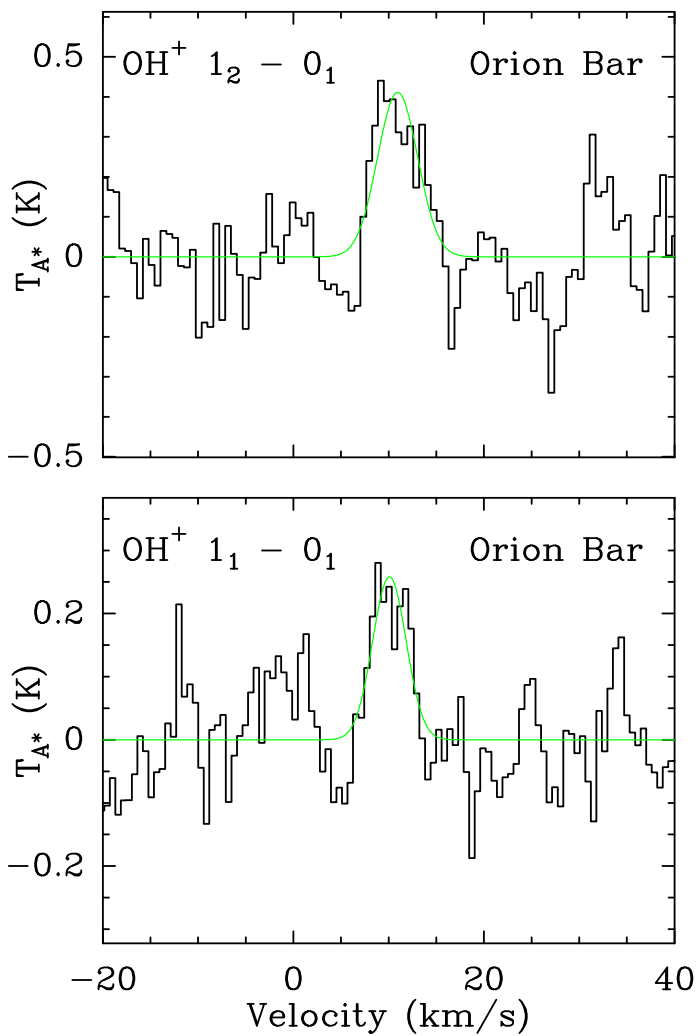}
\caption{Line profiles of the \ohp\ $1_2$--$0_1$ (top) and $1_1$--$0_1$ (bottom) transitions, observed with HIFI toward the \new{\cop\ peak of the} Orion Bar, \new{after smoothing to 0.6~\kms\ resolution.}}
\label{f:profi}
\end{figure} 

\begin{table}
\caption{Line parameters measured from the HIFI spectra.}
\label{t:pars}
\begin{tabular}{ccccccc}
\hline \hline
\noalign{\smallskip}
Line & $\int$ \tas\ \dv\ & \vlsr\ & \dv\  & rms  \\
         & \kkms\                 &  \kms\  & \kms\ & mK  \\ 
\noalign{\smallskip}
\hline
\noalign{\smallskip}
\ohp\ $1_0$--$0_1$ & $<$0.80 & ... & ... &  \phantom{1}58 \\
\ohp\ $1_2$--$0_1$ & 2.3(3) & 10.9(3) & 5.1(6) & 133 \\ 
\ohp\ $1_1$--$0_1$ & 1.2(1) & 10.1(3) & 4.3(5) & \phantom{1}88 \\
\hhop\ $1_{10}$--$1_{01}$ & $<$0.18 & ... & ... &  \phantom{1}20 \\
\hhop\ $1_{11}$--$0_{00}$ & $<$1.90 & ... & ... & 112 \\
\hhhop\ $0_0^-$--$1_0^+$ & $<$1.17 & ... & ... &  \phantom{1}78 \\
\hhhop\ $1_1^-$--$1_1^+$ & $<$2.27 & ... & ... &  \phantom{1}90 \\
\noalign{\smallskip}
\hline
\noalign{\smallskip}
\end{tabular}
\tablefoot{Numbers in parentheses are error bars in units of the last decimal. Noise levels in the last column are for a channel width of 1~MHz. Upper limits in column 2 are for \dv\ = 4.3\,\kms.
}
\end{table}

\begin{figure}[t]
\centering
\includegraphics[width=6cm,angle=-90]{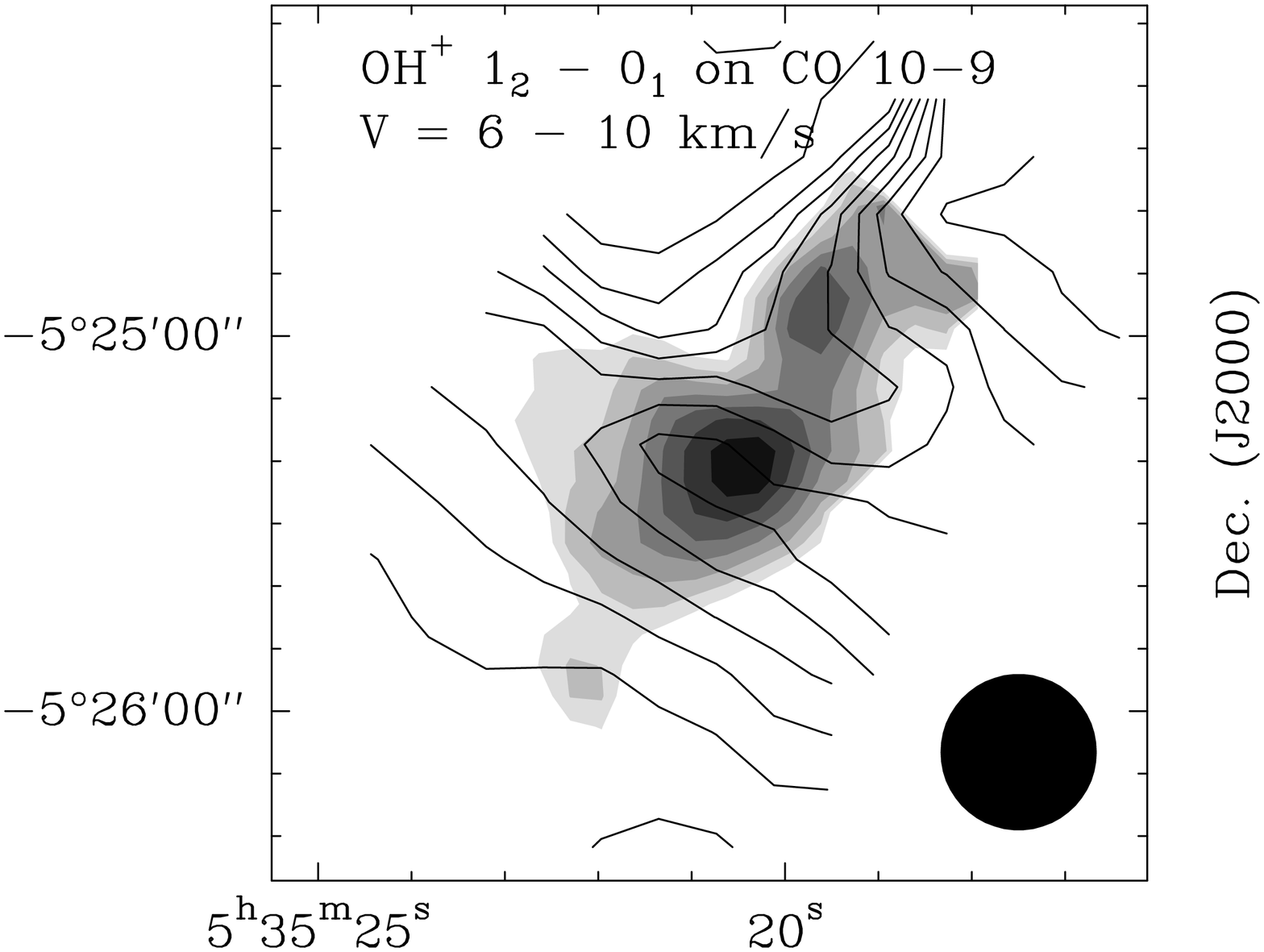}
\bigskip
\includegraphics[width=8cm,angle=0]{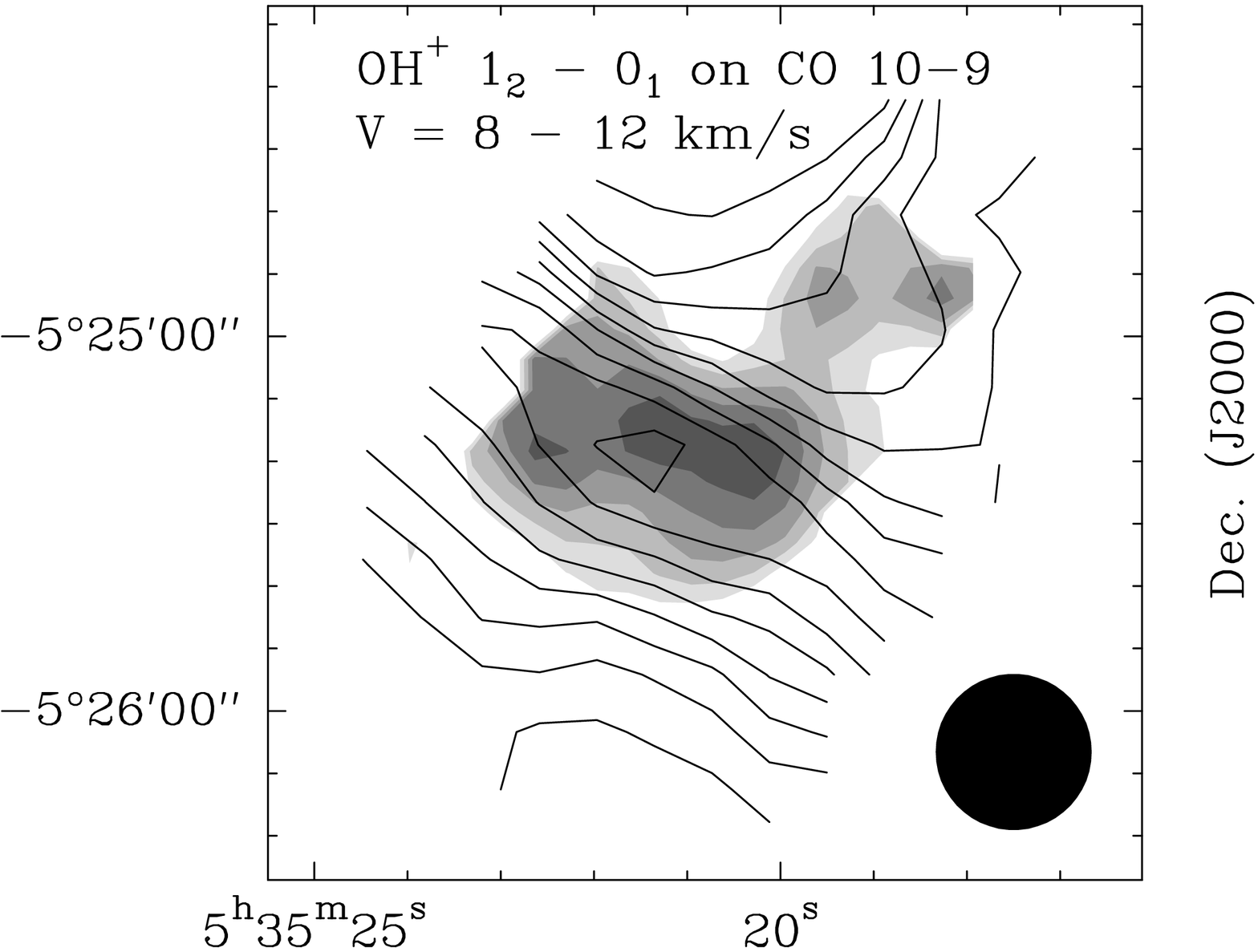}
\smallskip
\includegraphics[width=6.5cm,angle=-90]{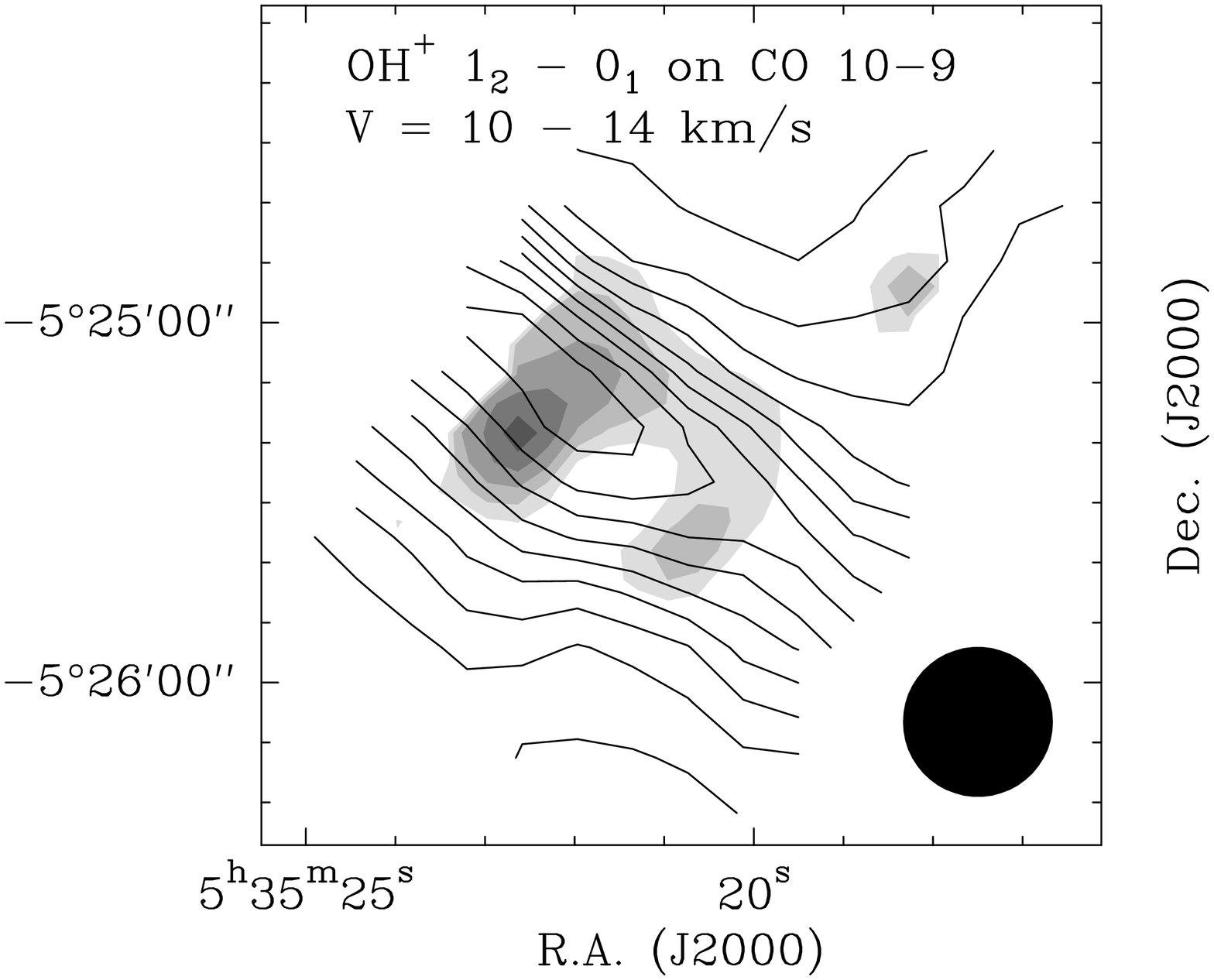}
\caption{Spatial distribution of the \ohp\ $1_2$--$0_1$ emission, integrated between \vlsr\ = +6 and +10 \kms\ (top), between +8 and +12 \kms\ (middle), and between +10 and +14 \kms\ (bottom). Greyscale levels start at 0.6 K\,\kms\ and increase by 0.2 K\,\kms. Contours of CO 10--9 emission are at 5, 15, ...95\% of the peak intensity in the respective velocity channel. \newest{Vibrationally excited \hh\ peaks near the North-West edge of the CO emission (cf. Fig. 1 of Nagy et al 2013)}.}
\label{f:maps}
\end{figure} 

\subsection{Line profiles}
\label{ss:prof}

Figure~\ref{f:profi} presents the line profiles of the \ohp\ lines at 971 and 1033 GHz, as observed with HIFI toward the Orion Bar.
Although hints of \ohp\ and \hhop\ emission have been seen toward the high-mass protostar W3 IRS5 \citepads{2010A&A...521L..35B}, this is the first time that a pure emission profile is seen toward a Galactic source.
As the line profiles appear single-peaked at our sensitivity and spectral resolution, we have fitted a Gaussian model to extract the line parameters.
Table~\ref{t:pars} presents the results of these Gaussian fits, and also reports our upper limits to other lines of \ohp, \hhop\ and \hhhop\ from our spectral survey of the Orion Bar.
We have also searched for the O$^+$ excited fine structure ($^2$D $J$ = 5/2 $\to$ 3/2) line near 593.7 GHz \citepads{2004ApJ...610..813B,2004ApJ...606..605S}, leading to an upper limit on \tas\ of 30~mK rms per 0.5 \kms\ channel.
No continuum emission is detected in the spectra, down to upper limits ranging from $T_{mb}^{SSB}$ = 0.1\,K at 607~GHz to 0.3\,K at 1655~GHz, which is consistent with the SPIRE measurements of \citetads{2012A&A...541A..19A}.

The difference between the measured central velocities and FWHM widths of the two detected \ohp\ lines is mostly due to hyperfine blending of the $1_2$--$0_1$ line. 
We thus adopt the measured parameters of the $1_1$--$0_1$ line as the best estimate of the central velocity and FWHM width of the \ohp\ emitting gas in the Orion Bar.
While the central velocity of this line of 10.1\,\kms\ is in good agreement with the value of 10.0$\pm$0.2 \kms\ from ground-based observations of other molecular species toward this source, its width of 4.3\,\kms\ is much larger than the value of 1.7$\pm$0.3 \kms\ measured for the dense gas deep inside ($A_V \gtsim 1$) the Bar (Hogerheijde et al \citeyearads{1995A&A...294..792H}, Leurini et al \citeyearads{2006A&A...454L..47L}). 
On the other hand, its width is smaller than the value of $\approx$5~\kms\ measured in species such as HF which trace the surface ($A_V \ltsim 0.1$) of the PDR \citepads{2012A&A...537L..10V}, and more similar to the width of the [C II] line \citep{2013A&A...550A..57O}.
Based on the line width, the \ohp\ emission thus seems to originate from close to the PDR surface ($0.3 < A_V < 0.5$) where \chp\ and \shp\ peak as well in the chemical model of \citetads{2013A&A...550A..96N}.
\new{In contrast, the non-hydride reactive molecular ions \cop, SO$^+$ and HOC$^+$ have smaller line widths \citepads[2--3\,\kms,]{2003A&A...406..899F}. These species have a different formation channel and thus a different sensitivity to the abundances of atomic H and electrons.}

\subsection{Spatial distribution}
\label{ss:map}

Figure~\ref{f:maps} presents our maps of the \ohp\ 971 GHz line emission, as observed with HIFI toward the Orion Bar, integrated over three velocity ranges, shown in greyscale.
The contours show the CO 10--9 line \new{emission}, integrated over the same velocity ranges.
The \ohp\ emission around the central velocity (middle panel) is seen to be extended over at least an arc minute (25,000~AU or 0.12~pc) on the sky, and to roughly follow the structure of the Bar seen in CO 10--9 and other molecular tracers \citepads[e.g.,]{2009A&A...498..161V}. 
The emission from the Bar is concentrated in two clumps: one peaking near the map center at \vlsr\ $\approx$10 \kms, and another peaking 20--25$''$ to the East near \vlsr\ = 12 \kms, which is seen most pronounced in the figure's bottom panel.
\new{This second peak lies $\approx$10$''$ South-East of the CO 10--9 peak at the same velocity, which coincides with H$^{13}$CN clumps 2 and 3 from \citetads{2003ApJ...597L.145L}. Since \ohp\ is not expected deeper into the PDR than CO and HCN, we suggest that the location of the second \ohp\ clump marks a deviation from pure plane-parallel geometry.}

The map of low-velocity \ohp\ in the top panel of Figure~\ref{f:maps} shows a third clump which lies to the North-West of the Bar, and is connected to it by a bridge of fainter \ohp\ emission. 
This `perpendicular' emission is also seen in CO 10--9 and other tracers of the Bar surface such as \new{OH 119\,$\mu$m \citepads{2011A&A...530L..16G}}, [CII] 158\,$\mu$m \citepads{2013A&A...550A..57O}, and \chp\ 3--2 \citepads{2013A&A...550A..96N}. 
The feature corresponds to the Southern tip of the Orion Ridge facing the Trapezium cluster, which confines the HII region, as visible in large-scale maps of CN line emission \citepads{1998A&A...329.1097R} and \thco\  3--2 emission (Lis \& Schilke \citeyearads{2003ApJ...597L.145L}; Melnick et al \citeyearads{2012ApJ...752...26M}).
From multi-line CN observations, \citetads{2001ApJ...559..985R} derive \hh\ densities as high as 10$^6$\,\ccm\ for this ridge, which may be favourable to excite \ohp\ line emission. 
However, this density is probably an overestimate, as collisions of CN with electrons were not taken into account, which are known to be important for CN \citepads{1991ApJ...369L...9B,2013MNRAS.435.3541H}.
Furthermore, we note that the \ohp\ line is not detected toward the Orion~S clump further up the Orion Ridge, possibly because emission and absorption from different layers cancel out each other.

\begin{table}
\caption{Optically thin estimates of \hnop\ column densities \newest{toward the \cop\ peak of the Orion Bar}, in units of 10$^{12}$ \scm, as a function of assumed excitation temperature.}
\label{t:cold}
\begin{tabular}{cccc}
\hline \hline
\noalign{\smallskip}
\txc\ (K) & \ohp\ & \hhop\ & \hhhop\   \\
\noalign{\smallskip}
\hline
\noalign{\smallskip}
10   & 82.1 & $<$2.85 & $<$7.21 \\
20   & 8.61 & $<$0.21 & $<$0.64 \\
40   & 4.01 & $<$0.10 & $<$0.27 \\
80   & 3.90 & $<$0.13 & $<$0.32 \\
160 & 5.45 & $<$0.25 & $<$0.64 \\
\noalign{\smallskip}
\hline
\noalign{\smallskip}
\end{tabular}
\tablefoot{The limits for \hhop\ and \hhhop\ are based on the 607~and 985~GHz lines; limits from the other lines are significantly higher.}
\end{table}

\begin{figure}[t]
\centering
\includegraphics[width=6cm,angle=-90]{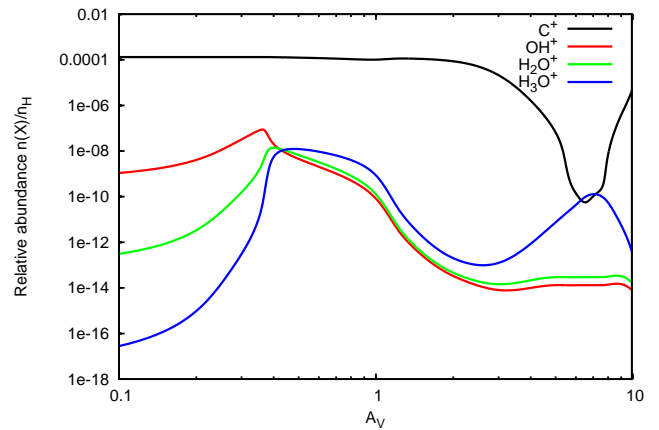}
\caption{Abundances of \ohp, \hhop, \hhhop, \newest{and \cp}, predicted by the Meudon PDR model as a function of visual extinction for a radiation field of $\chi$=10,000 \newest{in units of the local interstellar radiation field according to \citetads{1978ApJS...36..595D}} and a pressure of $P$=10$^8$ K\,cm$^{-3}$, as applicable to the Orion Bar.}
\label{f:pdr}
\end{figure} 

\subsection{Column densities}
\label{ss:cold}

The variations in \ohp\ emission level by factors of 2--3 across the maps in Figure~\ref{f:maps} likely correspond to variations in the total column density or the \ohp\ abundance with position, although excitation conditions (volume density, kinetic temperature) may also play a role.
\oldest{Since the emission does not appear to be strongly peaked,}
We focus in the following analysis on the \cop\ peak where we have limits on the other \hnop\ ions.
To estimate the column densities of these ions, we use the line fluxes from Table~\ref{t:pars} and apply a main beam efficiency of 76\% at 607~GHz, 70\% at 1655~GHz, and 74\% at our other line frequencies \citepads{2012A&A...537A..17R}. The column density $N$ depends on the excitation temperature $T_{\rm ex}$ through

$$ N_{\rm tot} = \frac{8\pi k \nu^2 }{hc^3} \frac{Q(T_{\rm ex})}{g_u A_{\rm ul}} e^{E_u / kT_{\rm ex}} \int T_{\rm mb} dV$$

where $\nu$ is the line frequency, $Q$ is the partition function, $g_u$ is the upper state degeneracy, and $A_{ul}$ is the spontaneous decay rate. 
This equation assumes optically thin emission and negligible background radiation ($T_{\rm bg} << T_{\rm ex}$), but does not assume the Rayleigh-Jeans limit ($h\nu << kT_{\rm ex}$) because our measurements are at high frequency.
The optically thin assumption is reasonable given the low expected abundances of the \hnop\ ions, and neglecting background radiation is justified given the low level of continuum radiation in our observations.
Adopting a background brightness temperature of 9~K increases the column density estimates by 5\% for \txc\ = 20~K, and by $<$1\% for higher values of \txc.

To evaluate the above expression for the column density, we use the spectroscopic parameters of the lines in Table~\ref{t:lines}. 
Table~\ref{t:cold} presents estimates of the column densities of \ohp, \hhop\ and \hhhop\ for values of \txc\ between 10~and 160~K, which is the expected range for the Orion Bar. 
If the excitation of \hnop\ is close to LTE, \txc\ would be close to the kinetic temperature of the gas, which ranges from $\approx$85~K for the dense gas \citepads{1995A&A...294..792H} to $\approx$150~K near the cloud surface \citepads{2003A&A...408..231B,2011A&A...530L..16G}. 
If collisional excitation of the lines cannot compete with their radiative decay, \txc\ will drop below \tkin, while \txc\ may exceed \tkin\ if radiative or chemical pumping plays a role. 
Section~\ref{s:nonlte} discusses these processes in more detail.
For now, the table shows that the derived column densities of \hnop\ vary by factors of 2--3 for \txc\ $\gtsim$20~K, but that the estimate increases by an order of magnitude if \txc\ is as low as 10~K.

The \ohp\ column density in Table~\ref{t:cold} for \txc\ = 10~K is comparable to the values toward other Galactic sources, while the estimates for \txc\ $\gtsim$20~K are $\sim$10$\times$ lower. 
Absorption line studies indicate $N$(\ohp) values between a few 10$^{13}$ and a few 10$^{14}$~\scm, both for the diffuse foreground clouds toward G10.6, W49N, W51 and OMC-2 \citepads{2010A&A...518L.110G,2010A&A...521L..10N,2012ApJ...758...83I,2013A&A...549A.114L} and the dense gas around the protostars AFGL 2591, W3 IRS5, and Orion-KL \citepads{2010A&A...521L..44B,2010A&A...521L..35B,2010A&A...521L..47G}.
Our \ohp/\hhop\ ratio of $>$40 and \ohp/\hhhop\ ratio of $>$15 are larger than in previous observations of diffuse clouds, suggesting an origin of the observed emission in very diffuse low-extinction layers of the PDR.

\section{PDR models}
\label{s:pdr}

\begin{figure}[t]
\centering
\includegraphics[width=6cm,angle=0]{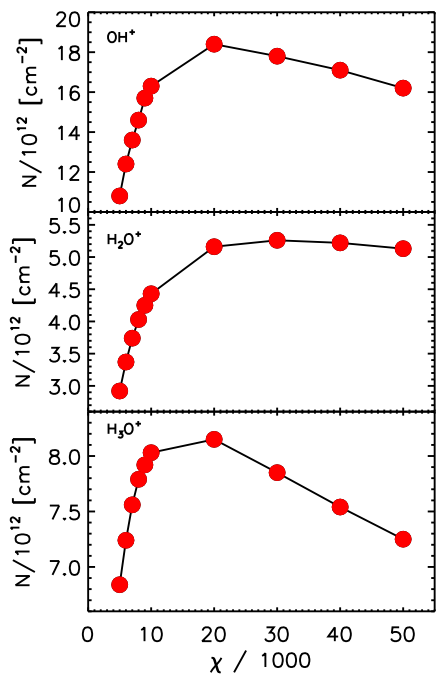}
\caption{\newest{Face-on} column densities of \ohp, \hhop\ and \hhhop, predicted by the Meudon PDR model as a function of radiation field (in Draine units) for a pressure of $P$=10$^8$ K\,cm$^{-3}$, as applicable to the Orion Bar. \new{The integrated $A_{\rm V}$ of the model is 10 mag.}}
\label{f:chi}
\end{figure} 

\begin{table}[t]
\begin{minipage}[t]{\linewidth}\centering
\caption{Rate coefficients of the main formation and destruction paths of H$_n$O$^+$. }
\label{table:rates}
\renewcommand{\footnoterule}{}
\begin{tabular}{lll}
\hline \hline
\noalign{\smallskip}
Reaction& $k$ (cm$^{3}$\,\ps) & $\Delta T$ (K) \\
\noalign{\smallskip}
\hline
\noalign{\smallskip}
$k_1$& $1.70\times10^{-9}$&                             10--41000\\ 
$k_3$& $2.10\times10^{-9}$&                             10--41000\\
$k_4$& $1.01\times10^{-9}$&                             10--41000\\
$k_5$& $6.40\times10^{-10}$&                            10--41000\\ 
$k_7$& $6.30\times10^{-9}~(T/300)^{-0.48}$&             -- \tablefootmark{a} \\
$k_9$& $3.05\times10^{-7}~(T/300)^{-0.50}$&             10--1000\\
$k_{10}$& $8.60\times10^{-8}~(T/300)^{-0.50}$&          10--1000\\
\noalign{\smallskip}
\hline
\noalign{\smallskip}
Reaction& $k$ (\ps) & $\Delta T$ (K) \\
\noalign{\smallskip}
\hline
\noalign{\smallskip}
$k_2$& $1.60\times10^{-12}~\exp{(-3.1 A_{\rm{V}})}$&    10--41000\\
$k_6$& $1.00\times10^{-9}~\exp{(-1.7 A_{\rm{V}})}$&    -- \tablefootmark{b} \\
$k_8$& $1.10\times10^{-11}~\exp{(-3.5 A_{\rm{V}})}$&    10--41000\\
\noalign{\smallskip}
\hline
\noalign{\smallskip}
\end{tabular}
\end{minipage}
\tablefoot{Rates are based on the UMIST database \citepads{2007A&A...466.1197W} unless otherwise noted. The photoreaction rates at the bottom are given for a standard Draine radiation field. The last column gives the temperature range over which the $k$-values are valid.}
\tablefoottext{a}{Rate from the OSU database (version March 2008).} \\
\tablefoottext{b}{Assumed value.}
\end{table}

\new{To understand the chemistry of \hnop\ in more detail, we model the Orion Bar with version 1.4.4 of the Meudon PDR model \citepads{2006ApJS..164..506L,2007A&A...467....1G,2008A&A...485..127G}. 
We describe the object with an isobaric model, where the pressure is kept constant and the program solves for the thermal and chemical balance as a function of depth.
Isobaric models are the simplest description of gas in steady-state, stationary molecular clouds where gravitation is negligible.
The model includes a gas-phase chemical network based on the UMIST database, as well as \hh\ formation on grains and neutralization of ions on grains and PAHs.}

Figure~\ref{f:pdr} shows the predicted abundances of \ohp, \hhop\ and \hhhop\ as a function of visual extinction, for a radiation field of $\chi$=10,000 Draine (\citeyearads{1978ApJS...36..595D}) units ($\chi_0$ = \pow{2.7}{-3} erg\,\ps\,\scm), \newest{a cosmic-ray ionization rate of \pow{2}{-16}\,\ps, } and a pressure of $P$=10$^8$ K\,cm$^{-3}$.
\new{The radiation field is taken from \citet{2011A&A...530L..16G}}
and the pressure is chosen to match the intensities of the \chp\ $J$ = 1--0 to 6--5 transitions observed toward the Orion Bar CO$^+$ peak with Herschel HIFI and PACS \citepads{2013A&A...550A..96N}. 
For these parameters, the total gas density at the depth where the H$_n$O$^+$ abundances peak ($A_{\rm{V}}\sim0.3-1.0$) is in the range between $5.6\times10^4$ and $1.8\times10^5$ cm$^{-3}$.
\new{The temperatures at these depths in the model (Table~\ref{table:rates_meudon}) agree well with estimates for the PDR surface from lines of \hh\ \citepads[400--700\,K,]{2005ApJ...630..368A} and H \citepads[$\sim$550\,K,]{2013ApJ...762..101V}.}
The column densities of our observed species, integrated up to $A_V$=1~mag, are $N$(\ohp) = \pow{1.6}{13}~\scm, $N$(\hhop) = \pow{4.4}{12}~\scm, and $N$(\hhhop) = \pow{7.5}{12}~\scm. 
When integrating up to $A_V$=10~mag, the values for \ohp\ and \hhop\ remain the same to 1\%, and $N$(\hhhop) increases by just 6\% which is insignificant. 
These predictions are consistent with the estimates in Table~\ref{t:cold} for \txc\ = 10--20~K, except that the observed \ohp/\hhop\ ratio is larger than in the models.
Note that the predictions correspond to a face-on model; to compare to the observations, they must be multiplied by a factor of 1/sin~$i$ $\approx$4, where $i\approx15^\circ$ for the Orion Bar \newer{\citepads{2013A&A...550A..96N}}.

As shown in Figure~\ref{f:pdr}, H$_n$O$^+$ abundances in the high UV illumination Orion Bar PDR peak near the cloud surface, at depths of $A_{\rm{V}}\sim0.3-0.4$, and decrease rapidly beyond $A_{\rm{V}}\sim1$. 
\newest{Given the clumpy structure of the Orion Bar, this narrow range in $A_{\rm{V}}$ is spatially rather extended, as indicated by the maps in Fig.~\ref{f:maps}.}
To understand the chemistry of \hnop\ in the Orion Bar, we study the main formation and destruction channels of these ions at depths of $A_{\rm{V}}$ = 0.3, 0.4 and 1.0 mag.
The corresponding reaction coefficients and rates are listed in Tables~\ref{table:rates} and~\ref{table:rates_meudon}.
At a depth of $A_{\rm{V}}$ = 0.3 -- 0.4, the dominant pathways for OH$^+$ formation are:

\begin{equation}
\label{k1}
\mathrm{H_2 + O^+ \xrightarrow{k_1} OH^+ + H}
\end{equation}

\begin{equation}
\label{k2}
\mathrm{OH + h\nu \xrightarrow{k_2} OH^+ + e^-}
\end{equation}

\begin{equation}
\label{k3}
\mathrm{OH + H^+ \xrightarrow{k_3} OH^+ + H}
\end{equation}

The dominant OH$^+$ formation path is via H$_2$ and O$^+$, which accounts for \new{70}\% of the total OH$^+$ production at $A_{\rm{V}} = 0.3$ and \new{85}\% at $A_{\rm{V}} = 0.4$. Photoionization of OH accounts for \new{18}\% of OH$^+$ formed at $A_{\rm{V}} = 0.3$ and \new{12}\% at $A_{\rm{V}} = 0.4$. 
Charge exchange between OH and H$^+$ is mostly significant at $A_{\rm{V}}\sim0.3$, resulting in \new{12}\% of OH$^+$ production which decreases to \new{3}\% at $A_{\rm{V}}\sim0.4$. 
This reaction is more important in X-ray dominated regions (XDRs), which is a key difference between such regions and high-illumination PDRs \citep{2010A&A...518L..42V}.
At a depth of $A_{\rm{V}}\sim1$, the path via H$_2$ and O$^+$ is still the dominant OH$^+$ formation path (72.1\%). 
However, the role of OH photoionization increases, as it accounts for producing \new{27}\% of OH$^+$ at this depth.

Once OH$^+$ is formed, H$_2$O$^+$ and H$_3$O$^+$ formation rapidly follows via similar reactions with H$_2$: 

\begin{equation}
\label{k4}
\mathrm{OH^+ + H_2 \xrightarrow{k_4} H_2O^+ + H}
\end{equation}

\begin{equation}
\label{k5}
\mathrm{H_2O^+ + H_2 \xrightarrow{k_5} H_3O^+ + H}
\end{equation}

At a depth of $A_{\rm{V}} = 0.3$, \new{100\% of} H$_2$O$^+$ forms from OH$^+$ in the reaction with H$_2$. 
At a depth of $A_{\rm{V}} = 0.4$, as the H$_3$O$^+$ abundance is increased compared to that at $A_{\rm{V}}$ = 0.3, a secondary reaction apart from the reaction from H$_2$ (83\%) produces \new{17}\% of the H$_2$O$^+$ formed at this depth:

\begin{equation}
\label{k6}
\mathrm{H_3O^+ + h\nu \xrightarrow{k_6} H_2O^+ + H}
\end{equation}

At a depth of $A_{\rm{V}}$ = 1.0, the role of H$_2$O$^+$ production from H$_3$O$^+$ decreases to \new{9}\%, compared to the dominant reaction from H$_2$ (\new{88}\%). 
At each of these depths in the model,100\% of H$_3$O$^+$ is produced in the reaction of H$_2$O$^+$ with H$_2$. 

The destruction of OH$^+$ is mainly via reactions with electrons, photons and H$_2$. 
The reaction with H$_2$ (Eqn. \ref{k4}) is the most important H$_2$O$^+$ formation path, as mentioned above. 
It is also the most important destruction path for OH$^+$ at depths of $A_{\rm{V}}$ = 0.4 and beyond.
At smaller depths, both dissociative recombination with electrons and UV photodissociation contribute significantly to the OH$^+$ destruction: at $A_{\rm{V}}\sim~0.3$, 

\begin{equation}
\label{k7}
\mathrm{OH^+ + e^- \xrightarrow{k_7} H + O}
\end{equation}

\noindent
contributes \new{52}\%, and

\begin{equation}
\label{k8}
\mathrm{OH^+ + h\nu \xrightarrow{k_8} H + O^+}
\end{equation}

\noindent
contributes \new{18}\%.
Destruction of H$_2$O$^+$ occurs mainly via dissociative recombination with electrons at low depths ($A_{\rm{V}}$ = 0.3 -- 0.4):

\begin{equation}
\label{k9}
\mathrm{H_2O^+ + e^- \xrightarrow{k_9} H + O + H}
\end{equation}

\begin{equation}
\label{k10}
\mathrm{H_2O^+ + e^- \xrightarrow{k_{10}} H + OH}
\end{equation}

Reaction \ref{k9} accounts for \new{71}\% of H$_2$O$^+$ destruction at $A_{\rm{V}}$ = 0.3, while reaction \ref{k10} contributes \new{20}\%. 
At a depth of $A_{\rm{V}}$ = 0.4, reactions \ref{k9} and \ref{k10} contribute \new{44} and \new{12}\% (respectively) to the H$_2$O$^+$ destruction.
\new{As \hhop\ does not have photodissociation channels longward of 13.6 eV, the model does not include this process} \newer{\citepads{2012ApJ...754..105H}}.
Destruction by H$_2$, which controls H$_3$O$^+$ production is only significant at $A_{\rm{V}}\sim1$ (\new{87}\%). 
Therefore, the high electron density and UV radiation field \new{in the Orion Bar may} explain the H$_2$O$^+$ and H$_3$O$^+$ non-detections, as a significant fraction of OH$^+$ and H$_2$O$^+$ is directly destroyed by UV photons or by recombination with electrons, limiting the H$_2$O$^+$ and H$_3$O$^+$ formation.
Our non-detections of \hhop\ and \hhhop\ are thus consistent with an origin of the \ohp\ emission in diffuse gas where \cp\ is abundant.

Alternatively, our observed \hhop/\ohp\ ratio may be due to a low molecular fraction (\hh/H ratio) in the gas probed by our observations.
Gerin et al. (2010) derive an analytic expression for the OH$^+$/H$_2$O$^+$ abundance ratio as a function of the gas temperature, electron density, and H$_2$ volume density:
\begin{equation}
\nonumber
n({\rm{OH^+}})/n({\rm{H_2O^+}})=0.64 + 430 \times (T/300)^{-0.5} \times [n({\rm{e}}^-)/n({\rm{H}}_2)]
\end{equation}
As most OH$^+$ forms in the outermost layers of the PDR ($A_{\rm{V}}<0.4$) at very low molecular fractions (0.01\%-0.4\% for $A_{\rm{V}}=0.3$ and $A_{\rm{V}}=0.4$, respectively), this formula is consistent with the observed abundance ratio of $N$(OH$^+$)/$N$(H$_2$O$^+$)$>$2. 
This suggests that, apart from the effect of the high electron density mentioned above, the low molecular fraction in the surface of the PDR also contributes to the observed OH$^+$/H$_2$O$^+$ column density ratios.
We conclude that most of our observed \newer{\ohp}\ emission originates at extinctions below $A_V = 0.4$.

Fig.~\ref{f:pdr} shows a second peak in the H$_3$O$^+$ abundance at a depth of $A_{\rm{V}}\sim7-8$. 
This second peak is expected in the abundance of H$_n$O$^+$ in interstellar clouds with a large range of physical conditions \citepads[e.g.,]{2012ApJ...754..105H} and does not significantly contribute to the total H$_3$O$^+$ column density. 
The sequence of H$_n$O$^+$ formation at this depth is initiated by the cosmic-ray ionization of H$_2$, followed by a reaction of H$_2^+$ with H$_2$ producing H$_3^+$. 
At this depth, OH$^+$  is formed from H$_3^+$, whose abundance is increased due to the lower electron abundance. 
The second abundance peak disappears toward higher $A_V$ as oxygen freezes out on the grain surfaces as water.  

Fig.~\ref{f:chi} shows the effect of changing the radiation field on the \hnop\ column densities. 
The model corresponds to $P$=10$^8$ cm$^{-3}$\,K and the figure considers radiation fields between $\chi$=5000 and 50000 Draine units. 
The predicted column densities of the \hnop\ species are seen to peak for $\chi$=20,000--30,000 and to drop for higher radiation fields.
\new{The \hnop\ column densities in our model are similar to those by \citetads{2012ApJ...754..105H} under Orion Bar conditions (their Figs.~8 and~9).}

\section{Excitation of \ohp}
\label{s:nonlte}

\begin{table}[t]
\begin{minipage}[!h]{\linewidth}\centering
\caption{Rates (in cm$^{-3}$ s$^{-1}$) corresponding to the main formation and destruction paths of H$_n$O$^+$ in the model at $A_{\rm{V}}=0.3$, 0.4, and 1.0.}
\label{table:rates_meudon}
\renewcommand{\footnoterule}{}
\begin{tabular}{lccc}
\hline \hline
\noalign{\smallskip}
$A_{\rm{V}}$ &   0.3 &  0.4 &  1.0 \\
$T_{\rm gas}$  & 1783 K &  1013 K  &  569 K \\
$n_{\rm gas}$ & \pow{5.6}{4}\,\ccm & \pow{9.9}{4}\,\ccm & \pow{1.8}{5}\,\ccm \\
$n$(\hh) & \pow{1.9}{1}\,\ccm & \pow{3.9}{3}\,\ccm & \pow{1.0}{5}\,\ccm \\
\noalign{\smallskip}
\hline
\noalign{\smallskip}
Reaction& \multicolumn{3}{c}{Rate}\\
\noalign{\smallskip}
\hline
\noalign{\smallskip}
$k_1$&    $1.31\times10^{-10}$&  $6.25\times10^{-9}$&  $1.53\times10^{-9}$\\ 
$k_2$&    $3.38\times10^{-11}$& $8.46\times10^{-10}$& $5.77\times10^{-10}$\\
$k_3$&    $2.15\times10^{-11}$& $2.25\times10^{-10}$& $1.22\times10^{-11}$\\
$k_4$&    $3.11\times10^{-11}$&  $7.14\times10^{-9}$&  $2.11\times10^{-9}$\\
$k_5$&    $9.52\times10^{-14}$&  $3.29\times10^{-9}$&  $2.08\times10^{-9}$\\ 
$k_6$&    $4.84\times10^{-14}$&  $1.43\times10^{-9}$&  $2.03\times10^{-10}$\\
$k_7$&    $9.75\times10^{-11}$&  $1.09\times10^{-10}$& $2.93\times10^{-12}$\\
$k_8$&    $3.27\times10^{-11}$&  $2.50\times10^{-11}$& $3.54\times10^{-14}$ \\
$k_9$&    $2.20\times10^{-11}$&  $3.76\times10^{-9}$&  $2.18\times10^{-10}$\\
$k_{10}$& $6.21\times10^{-12}$&  $1.06\times10^{-9}$&  $6.15\times10^{-11}$\\
\noalign{\smallskip}
\hline
\end{tabular}
\end{minipage}
\end{table}

Our estimated column densities of \ohp, \hhop\ and \hhhop\ are in reasonable agreement with the predictions from the Meudon PDR model, but the appearance of the lines in emission \newest{is unlike other Galactic sources observed in \ohp\ so far}. 
To understand this behaviour we perform a non-LTE analysis of the excitation of \ohp, which takes both reactive and inelastic collisions into account, as well as excitation by the background continuum radiation field. 

\subsection{Collisional and radiative excitation}
\label{ss:radex}

The calculations in Appendix~\ref{ss:crc} indicate inelastic electron collision rates for the \ohp\ 971 GHz line of $\approx$\pow{6}{-7} \ccps, which is $\sim$10$\times$ higher than the dissociative recombination rate at $T \approx$100~K (Table~\ref{table:rates}).
At this temperature, $h \nu / k T \approx 0.5$ for this transition, so that by detailed balance, the upward and downward rates differ by only a few percent.
For collisions with \hh, inelastic collision rates are not known, but it may be reasonable to assume that reactive collisions dominate. 
In contrast, the reaction of \ohp\ with H is endothermic, and inelastic collisions with H could influence the excitation of \ohp.
\new{Test calculations assuming a collisional rate coefficient of $10^{-10}$\,\ccps\ (cf. Andersson et al \citeyearads{2008ApJ...678.1042A} for the case of \cop) for all radiatively allowed transitions indicate that this effect is comparable with electron collisions.
\newest{Very recent calculations on the \ohp-He system confirm this estimate, at least to order of magnitude (F. Lique, priv. comm.).}
We therefore first calculate the excitation of \ohp\ assuming that inelastic collisions with H and electrons dominate, and consider the effect of reactive collisions with \hh\ in \S\ref{ss:form-dest}.}

We have used the non-LTE radiative transfer program Radex \citepads{2007A&A...468..627V} to calculate the excitation of \ohp\ in the Orion Bar, assuming steady-state conditions and using the inelastic collision rates from Appendix~\ref{ss:crc}.
Observations of C recombination lines toward the Orion Bar surface indicate hydrogen densities of \pow{5}{4}--\pow{2.5}{5} \ccm\ \citepads{1997ApJ...487L.171W}, which for all carbon in C$^+$ and C/H = \pow{1.4}{-4} translates into an electron density of $n$(e) $\approx$10 \ccm, as we have used before for HF \citepads{2012A&A...537L..10V}. 
We use an electron temperature of $T_e$ = 300 K, but our results are insensitive to variations in $T_e$ between 100 and 1000~K.
The adopted line width is the observed 4.3 \kms\ and for the background radiation field we adopt a \newest{modified blackbody} distribution with a dust temperature of $T_d$=50~K and a dust emissivity index of $\beta$=1.6, as found by \citet{2012A&A...541A..19A} for the interior of the Bar, so that $\tau_d$=0.21 at 971~GHz. 
The model is insensitive to the details of this radiation field; in particular, the results are unchanged when adopting $T_d$=70~K and $\beta$=1.2 as found by Arab et al for the Bar's surface.

The model predicts an excitation temperature of $\approx$10~K for the \ohp\ lines near 1~THz. 
The excitation is due to the combination of collisions and radiation: a model with the \newest{modified blackbody} replaced by the 2.73~K cosmic microwave background results in \txc\ $\approx$7~K.
The effect on the emerging line intensities is large, since the Planck function at 1~THz increases by a factor of $\sim$4 from \txc\ = 7~K to 9~K.

The calculated excitation temperature of $\approx$10~K is above the background radiation temperature of 9.2~K, which explains the appearance of the line in emission, but
the models require $N$(\ohp) $\approx$\pow{5}{14}\,\scm\ to match our observed 971~GHz line intensity. 
At this high column density, the \ohp\ lines become optically thick which drives the three $N$=1--0 lines to approximately the same peak brightness, inconsistent with our non-detection of the $1_0-0_1$ line of \ohp. 
We therefore regard this model as untenable, conclude that the observed line emission is optically thin, and search for a model with a higher excitation temperature in the next section.

\subsection{Effect of reactive collisions}
\label{ss:form-dest}

The above calculation indicates that \new{inelastic} collisions are sufficient to make the \ohp\ line appear in emission given the low background intensity of the Orion Bar. 
A fully self-consistent model should however include formation and destruction terms in the rate equations. 
We use Equation 12 from \citetads{2007A&A...468..627V}, where the number density of \ohp\ molecules (in \ccm) is the ratio of its formation rate (in \ccm\,\ps) and its destruction rate (in \ps). 
We adopt a destruction rate of \ohp\ in the Orion Bar of 10$^{-4}$~\ps, which is due to reactions with \hh, electron recombination, and photodissociation.
In contrast, the formation of \ohp\ through ion-molecule reactions should proceed at a rate at or below the Langevin rate of $\approx$10$^{-9}$~\ccps\ and is limited by the supply of hydrogen or oxygen ions.

Lacking state-to-state formation rates, we approximate the distribution of the newly-formed \ohp\ over its energy levels by a thermal distribution at a temperature $T_f$. 
The value of $T_f$ depends on the dominant formation route and is likely to be a significant fraction of its excess energy. 
The reactions of \op\ with \hh\ and of O with \hhhp\ have exothermicities of 0.55 and 0.66~eV \citepads{1984PhRvL..52.2084F,2000CPL...319..482M}, so $T_f$ should be in the range 2000--3000~K.

For \ohp\ formation rates between 10$^{-12}$ and 10$^{-9}$~\ccps\ and formation temperatures of 2000--3000~K, the model predicts excitation temperatures of $\approx$12~K for the \ohp\ lines near 1~THz. 
Compared to models with a negligible formation rate, the line brightness increases by a factor of $\approx$5, almost independent of $T_f$, for the same column density. 
The observed \newest{intensities of the $1_2--0_1$ and $1_1--0_1$ lines are} matched for $N$(\ohp) $\approx${\pow{1}{14}~\scm, which is considerably lower than for the steady-state excitation model.
The difference with the estimates in Table~\ref{t:cold} is the presence of background radiation.
\newest{This model predicts some emission in the $1_0--0_1$ line, but only at the 2$\sigma$ level of our observations.}
We conclude that formation pumping plays an important role for \ohp\ in the Orion Bar.

\section{Discussion}
\label{s:disc}

\subsection{Effect of X-ray ionization}
\label{ss:zeta}

\newest{Our derived \ohp\ column density of $\approx$\pow{1}{14}\,\scm\ is similar to that in diffuse clouds, as observed in absorption toward W49N and other sources (see references in \S\ref{ss:cold}), but $\sim$1.6$\times$ higher than the model prediction of \pow{6.4}{13}\,\scm\ in \S\ref{s:pdr}.}
The calculations in \S\ref{s:pdr} assume a cosmic-ray ionization rate of \pow{2}{-16}~s$^{-1}$, which is a typical value for diffuse interstellar clouds in the Solar neighbourhood \citepads{2012ApJ...745...91I} and an order of magnitude higher than the value for dense clouds \citepads{2000A&A...358L..79V}.
However, the actual ionization rate of the Orion Bar may be atypically high because of its proximity to the Trapezium stars.
The effects of X-rays from these stars on the chemistry are similar to those of cosmic rays \citep{2006ApJ...650L.103M}.
In particular, \citetads{2010A&A...521L..47G} estimate an ionization rate of \pow{3}{-15} \ps\ for the Orion KL region, dominated by X-rays from the star $\theta^1$C Ori.
Calculations using the Meudon code with an ionization rate of \pow{2}{-15}~s$^{-1}$ result in very similar \hnop\ column densities as before, though, presumably because of the high gas density in the Orion Bar.

\subsection{Possible ion sources}
\label{ss:ion-src}

\newest{Alternatively, the discrepancy between our observed and modeled column densities of \ohp\ may arise because the assumption of a stationary medium is inappropriate.}
Regarding the Orion Bar as a molecular cloud under external illumination and heating, the main source of \ohp\ may be the reaction of \hhhp\ with O, where the \hhhp\ is due to cosmic-ray ionization as well as leakage of UV photons from the ionized region. 
However, interaction of the PDR with the neighbouring photoionized nebula may supply \op\ ions which produce extra \ohp\ in their reaction with \hh.
The supply of \op\ ions may be due to leakage of hydrogen- and oxygen-ionizing photons, but also to advection of \hp\ and \op\ ions from the ionized nebula into the mostly neutral PDR, as in some planetary nebulae \citepads{1983IAUS..103...91B}\footnote{\newest{Very recently, \ohp\ emission has been detected towards the Helix nebula (Van Hoof et al, in prep).}}.
The concentration of \hp\ with depth into the PDR is then important, because charge transfer reactions (\hp\ + O $\leftrightarrow$ \op\ + H) will rapidly couple the \op/O ratio to the \hp/H ratio.
Observations of the [O{\sc III}] 88~\mic\ line toward the Orion Bar with Herschel/PACS show that ionized oxygen is widespread in the region (C. Joblin \& J. Goicoechea, priv. comm.); 
the emission extends well into the mostly-neutral gas traced by the [O{\sc I}] 63~\mic\ line (M. Gerin, priv. comm.), making the \op\ + \hh\ channel a likely source of \ohp\ \new{at low depths}. 
\new{Detailed comparison of the optical and near-IR (forbidden and/or permitted) lines of O and \op\ would be a stronger test of this scenario \citepads{2000A&A...364..301W,2011MNRAS.417..420M}}.
In addition, simulations of the full molecular + atomic + ionized gas in the Orion PDR with the Cloudy program \citepads{2013RMxAA..49..137F} would be useful to constrain the role of possible ion sources in the Orion Bar.

\subsection{Radiative pumping}
\label{ss:pump}

Besides changing the chemistry of \ohp, the strong infrared and ultraviolet radiation fields in the Orion Bar may change its excitation. 
Section~\ref{ss:radex} already showed that far-infrared continuum radiation raises the \ohp\ excitation temperature significantly above the level due to \new{inelastic} collisions alone.
In addition, mid-infrared pumping \newest{through the $v$=1--0 band at 3.38~\mic} may contribute if the radiative excitation rate of \ohp\ in the Bar exceeds the collisional excitation rate by electrons, which is $n$(e) $\times C_{\rm lu}$ = 10~\ccm\ $\times$\pow{6}{-7} \ccps\ = \pow{6}{-6}~\ps. 
The radiative rate is $B_{\rm lu} U_{\rm rad}$, the Einstein absorption coefficient times the radiative energy density, which is approximately $A_{\rm vib} \epsilon f / (e^{(h\nu / (kT_d))} -1)$, where $\epsilon$ is the dust emissivity, $f$ the dust filling factor, $A_{\rm vib}$ the spontaneous decay rate of 265~\ps, and $T_d$ the dust temperature. 
For the Orion Bar, the filling factor should be close to unity, and for the emissivity we assume 1 at the short wavelength of the fundamental vibrational band of \ohp.
Equating the radiative rate to the electron collision rate indicates a minimum temperature of $\approx$240~K, which is reasonable for the gas at the Bar's surface, but too much for the dust, as the PACS and SPIRE data show \citep{2012A&A...541A..19A}. 
Only PAHs and small grains would reach such high temperatures, but with very low opacities,
even though the \ohp\ vibrational fundamental is close to the PAH 3.3~\mic\ and aliphatic 3.4~\mic\ emission features. 

Besides continuum radiation from dust in the Bar itself, pumping by infrared starlight from the Trapezium may influence the excitation of \ohp.
The brightest of these stars is $\theta^1$C~Ori, which is 127$''$ away from our observing position \citepads{2007A&A...474..653V}. 
If both objects lie at the same distance from the Sun, the stellar continuum flux at the Orion Bar is 2.96 million times stronger than at the Earth. 
The stellar temperature of 37,000~K implies a radiative intensity of 0.2~Jy/nsr or \pow{2}{-15} erg\,s$^{-1}$\,cm$^{-2}$\,Hz$^{-1}$\,sr$^{-1}$. 
Setting this equal to the Planck function at a radiation temperature $T_R$, we obtain $T_R = 185$\,K at $\lambda$ = 3.38\,$\mu$m. 
The corresponding pumping rate in the \ohp\ vibrational fundamental is of order $A_{\rm vib} / \{ \exp(h\nu / kT_R) -1 \}$ = \pow{5.7}{-8}\,s$^{-1}$, which is much less than the collisional excitation rate by electrons. 

While infrared pumping does not seem play a role for \ohp\ in the Orion Bar, the absorption rate through electronic transitions in the near ultraviolet is rather higher. 
The radiative intensity of $\theta^1$C~Ori at the wavelength of the A$^3\Pi_i$ -- X$^3\Sigma^-$ $v$=0--0 band of \ohp\ at 27949\,cm$^{-1}$ (3577\,\AA) is $\approx$1.8\,Jy/nsr. 
The A-value for this band is \pow{8.01}{5}\,\ps\ \citepads{1981A&A....95..383D}, so that the absorption rate in this band alone is $\approx$\pow{3.3}{-6}\,\ps, which is only slightly less than the collisional excitation rate.

\subsection{Comparison with extragalactic systems}
\label{ss:xgal}

The Orion Bar is the first and so far only position within our Galaxy where lines of \ohp\ appear purely in emission, \new{although a mix of emission and absorption is seen in W3 IRS5 \citep{2010A&A...521L..35B} and NGC 3603 (Makai et al, in prep.) and possibly many more star-forming regions.}
Detections of extragalactic \ohp\ and \hhop\ emission have been made with Herschel-SPIRE by \citetads{2010A&A...518L..42V} toward the active nucleus of the galaxy Mrk~231, with Herschel-PACS toward the ultraluminous merger Arp~220 by \citetads{2011ApJ...743...94R} and with Herschel-SPIRE toward the Seyfert nucleus NGC~1068 by \citetads{2012ApJ...758..108S}. 
In addition, detections of extragalactic \hnop\ absorption exist toward M~82 using HIFI \citepads{2010A&A...521L...1W} and SPIRE \citepads{2012ApJ...753...70K}.
Recently, PACS observations of excited \ohp, \hhop\ and \hhhop\ toward NGC 4418 and Arp 220 have been discussed by \citetads{2013A&A...550A..25G}.
\old{We suspect that the nuclei where \hnop\ lines appear in emission have an enhanced electron density, far-infrared continuum, and/or ionizing (UV/X-ray) continuum.}
\new{The Orion Bar is special in our Galaxy for its large column density of warm, mostly-atomic gas and its weak far-infrared continuum; we suspect that extragalactic nuclei where \hnop\ lines appear in emission have similar conditions}. \newest{In addition, supernova remnants may contribute, as the recent detection of \ohp\ emission towards the Crab nebula suggests (Barlow et al, in prep.).}

\section{Conclusions}
\label{s:conc}

We have presented maps and spectra of \ohp\ line emission toward the Orion Bar, and limits on lines of \hhop\ and \hhhop. 
The \ohp\ line emission is extended over $\sim$1$'$ (=25,000~AU = 0.12~pc) and traces the Bar itself as well as the Southern tip of the Orion Ridge.
\new{Analysis of the chemistry and the excitation of \ohp\ } suggests an origin of the emission at a depth of $A_V$=0.3--0.5, similar to \chp\ and \shp. 
The \ohp\ column density of  $\approx$\pow{1.0}{14}~\scm, derived using a non-LTE model including both inelastic and reactive collisions and radiative pumping, is similar to that in previous absorption line studies, while our limits on the \ohp/\hhop\ and \ohp/\hhhop\ ratios are higher than seen before. 

Non-LTE models of the excitation of \ohp\ show that the unusual appearance of the \ohp\ lines in emission is the combined result of inelastic \old{electron} collisions, far-infrared radiative pumping by dust, and chemical pumping through the \op\ + \hh\ and O + \hhhp\ channels. 
The same conditions may apply to extragalactic sources of \hnop\ line emission.
In the future, high-resolution maps of Galactic and extragalactic \ohp\ line emission with ALMA will shed further light on the chemistry of this reactive ionic species.

Our observed \hnop\ column densities are qualitatively reproduced by a model of the Orion Bar (using the Meudon PDR code) using a radiation field of $\chi$=10$^4$~$\chi_0$ and a pressure of $P$=10$^8$~K\,\ccm\ as suggested by previous observations. 
Analysis of the main formation and destruction paths of the ions indicates that our high \ohp/\hhop\ and \ohp/\hhhop\ ratios are due to the high UV radiation field and electron density in the Orion Bar. 
Destruction of \ohp\ and \hhop\ by photodissociation and electron recombination limits the formation of \hhop\ and \hhhop.
\new{In addition, the low molecular fraction at the PDR surface limits the production of \hhop\ and \hhhop.}

Quantitatively, the Meudon PDR models underpredict the absolute \ohp\ column density by a factor of \newest{$\sim$1.6}.
To match the observed line intensity with an \ohp\ column density similar to that in the PDR model, the electron density would have to be $\approx$100\,\ccm, which is much higher than the PDR model predicts at the depth where the \hnop\ ions are abundant, as seen in Fig.~10 of \citetads{2013A&A...550A..96N}.
However, raising the pressure in the PDR model by $\approx$50\% would increase the predicted $N$(\ohp) to the value suggested by the \new{non-LTE} models. 
Such an increase is consistent with the \chp\ and \shp\ observations, and is also suggested by observations of high-$J$ CO lines with PACS (Joblin et al, in prep.).
Furthermore, the Meudon PDR model uses a scaling of the average interstellar radiation field, while realistic models should use direct observations of the dominant hot star $\theta^1$C~Ori for the H-ionizing part of the spectrum and the wavelength range where OH photoionization occurs, which could contribute significantly to the formation of \ohp.

\begin{acknowledgements}
{The authors thank S\'ebastien Bardeau and J\'er\^ome Pety (IRAM) for help with the data reduction, Franck le Petit (Paris-Meudon) for assistance with the PDR model calculations, Arturo Rodr\'{\i}guez-Franco (Madrid) for sending his CN map in electronic form, Xander Tielens (Leiden) for useful discussions, Inga Kamp (Groningen) for a careful reading of the manuscript,  Christine Joblin (Toulouse) \& Javier Goecoechea (Madrid) for sharing their PACS observations of the [O{\sc III}] line, \new{and the referee for a useful report}. 
AF acknowledges support by the Agence Nationale de la Recherche (ANR-HYDRIDES), contract ANR-12-BS05-0011-01.}

\par

{HIFI has been designed and built by a consortium of institutes and university departments from across Europe, Canada and the US under the leadership of SRON Netherlands Institute for Space Research, Groningen, The Netherlands with major contributions from Germany, France and the US. Consortium members are: Canada: CSA, U.Waterloo; France: CESR, LAB, LERMA, IRAM; Germany: KOSMA, MPIfR, MPS; Ireland, NUI Maynooth; Italy: ASI, IFSI-INAF, Arcetri-INAF; Netherlands: SRON, TUD; Poland: CAMK, CBK; Spain: Observatorio Astron\'omico Nacional (IGN), Centro de Astrobiolog\'{\i}a (CSIC-INTA); Sweden: Chalmers University of Technology - MC2, RSS \& GARD, Onsala Space Observatory, Swedish National Space Board, Stockholm University - Stockholm Observatory; Switzerland: ETH Z\"urich, FHNW; USA: Caltech, JPL, NHSC.}
\end{acknowledgements}

\bibliographystyle{aa}
\bibliography{hnop-ads}

\newpage

\appendix

\section{Inelastic collision rates for the \ohp -- \emin\ system}
\label{ss:crc}

 The electronic ground state symmetry of the radical OH$^+$ is
 $^3\Sigma^-$. Each rotational level $N$ is therefore split by the
 spin-rotation coupling between $N$ and the electronic spin $S=1$ so
 that each rotational level $N$ has two sub-levels given by $j=N\pm
 1$. In addition, owing to the non-zero nuclear spin of the hydrogen
 atom ($I$=1/2), each fine-structure level is further split into 2
 hyperfine levels $F=j\pm 1/2$. The rotational constant of OH$^+$ is 492.26~GHz.
 The fine-structure splitting is of the order of 60--90~GHz
 while the hyperfine splitting is less than 0.3~GHz.
 The dipole moment of OH$^+$ is 2.26~D 
 \citepads{1983JChPh..79..905W}.

Electron-impact hyperfine excitation rate coefficients for OH$^+$ were
computed using a three-step procedure: {\it i)} rotational excitation
rate coefficients for the dipolar ($\Delta N=1$) transitions were
first computed within the Coulomb-Born approximation; {\it ii)}
fine-structure excitation rate coefficients were then obtained from
the Coulomb-Born rotational rates using the (scaled)
infinite-order-sudden (IOS) approximation; and {\it iii)} hyperfine
excitation rate coefficients were finally obtained using the so-called
``statistical'' or ``proportional'' approach. The Coulomb-Born
approximation \citepads{1974PhRvA..10..788C} is expected to be accurate for
polar molecules with dipoles in excess of $\sim 2$~D because the
dipolar cross sections are entirely dominated by long-range effects
and cross sections for transitions with $\Delta N \geq 2$ are
significantly smaller 
\citepads{2001MNRAS.325..443F}. In practice,
Coulomb-Born cross sections were computed for collision energies below
2~eV and rate coefficients were deduced for temperatures ranging from
10 to 2000~K. The IOS approximation was employed to derive the
fine-structure rate coefficients in terms of the rotational rates for
excitation out of the lowest rotational level $N=0$. This IOS
formalism was first introduced by  
\citetads{1984JChPh..81.3892C} for
linear molecules with $^{2S+1}\Sigma$ symmetry. As the Coulomb-Born
rotational rates do not strictly obey the IOS factorization
formulae, however, the IOS fine-structure rate coefficients were
scaled, as recommended by 
\citetads{2012MNRAS.425..740F} (see their Eqs.~(8),
(10) and (13), where the quantum number $F$ should be replaced by
$S$). While in principle the IOS approximation should be also
applicable to obtain the hyperfine rate coefficients of OH$^+$, there
is to our knowledge no available factorization formula for a
$^3\Sigma$ molecule. Hyperfine rate coefficients were therefore
obtained from the fine-structure rates by assuming that they are
proportional to the degeneracy $(2F + 1)$ of the final hyperfine
level. We note that this simple statistical approach does not account
for the collisional propensity rule $\Delta F=\Delta j$. As shown by
\citetads{2012MNRAS.425..740F}, 
however, at low total optical depth ($\tau\leq
10$) the statistical approach is applicable because in this regime the
relative populations of each hyperfine component are close to the
statistical weights.

The above three-step procedure was applied to the first 49 levels of
OH$^+$, that is up to the level $(N, j, F)=(8, 8, 17/2)$ which lies 1689~K
above the ground state (0, 1, 3/2), resulting in 176
collisional transitions. A typical accuracy of 30\% is expected for
these rate coefficients, with the largest rate coefficients being of
the order of $3\times 10^{-6}$ cm$^3$s$^{-1}$.
The rates will be posted on the website of the LAMDA database \citepads{2005A&A...432..369S}\footnote{\tt http://home.strw.leidenuniv.nl/$\sim$moldata/}.

\end{document}